# PIC Simulations of Prompt GRB Emissions

E. Liang, K. Noguchi and S. Sugiyama

*Rice University, Houston, TX 77005-1892*

**Abstract**. We review PIC simulation results of GRB emissions for both the Poynting flux and shock scenarios.

## INTRODUCTION

An outstanding problem in γ-ray bursts (GRBs) is the particle acceleration and radiation mechanisms, which are so efficient that they convert most of the primary energy from the central engine into γ-rays. While there is increasing evidence that at

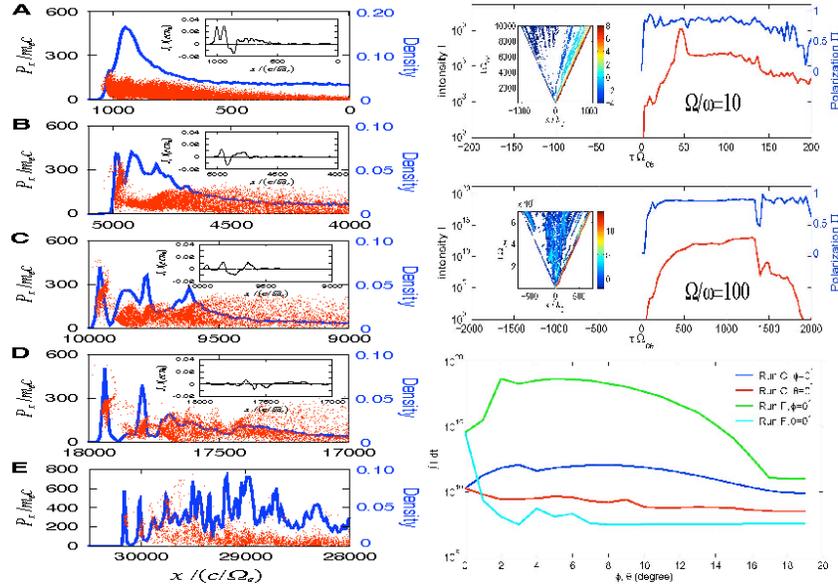

**FIGURE 1** (left) PIC simulated PF accelerated e+e- slab showing repeated bifurcation of pulse profile (curves) and momentum profiles (dots) which resemble GRB light curves and hardness evolution [7].
**FIGURE 2** (right) Using ray-tracing we compute the radiation intensity and polarization of a PF accelerated e+e- slab as functions of detector time (upper and middle panels) and angular fluence as function of view-angles (lower panel). Higher B field (middle panel) gives longer detected pulse [9].

least the classical long GRBs are related to the deaths of massive stars, there is yet little consensus on the primary energy source, or how it is converted into γ-rays. Two popular paradigms are relativistic hydrodynamic bulk flows (HBF) [1] or electromagnetic-dominated outflows (Poynting flux, PF) [2], driven by the formation of a new-born black hole or neutron star [3]. The former could be caused by disk accretion or a neutrino wind, while the latter could be caused by strongly magnetized accretion or ms magnetar wind. In both paradigms the difficulty is to find robust physical mechanisms which efficiently convert electromagnetic (EM) or bulk flow (BF) energy into the ultra-relativistic internal energy of a small number of nonthermal particles. A popular model of the HBF paradigm is the internal/external synchrotron shock model, where unsteady BF leads to dissipation via internal shocks in the prompt phase and snowplowing of the CSM/ISM leads to external shock emission in the afterglow phase. In shock models the popular mechanism is Fermi acceleration at the shock fronts followed by synchrotron or "jitter" radiation [4], though Comptonization [4] has also been invoked. Magnetic turbulence generated by Weibel instability is usually invoked in shock models [5]. In the PFA paradigm, particle acceleration and radiation are driven directly by large-scale ordered EM fields via collective plasma processes [2]. In both the HBF and PFA paradigms, the physics of relativistic particle acceleration and γ-radiation ultimately involve electromagnetic (EM) and plasma kinetic processes, which can only be correctly simulated using Particle-in-Cell (PIC) codes [6]. Such simulations have recently been pioneered by several groups [5,7]. Our group, working with LLNL and LANL, has developed and adapted the most advanced PIC codes for GRB simulation, including the first 3D PIC code with self-consistent radiation damping [8,9], and applied them to compute radiation outputs of both the HBF and PF scenarios [8,9]. Sample results are given in Fig.1-2.

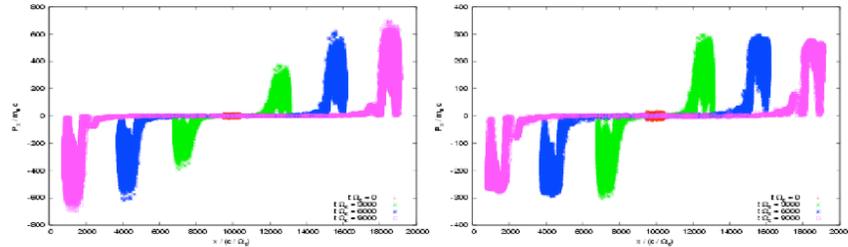

**FIGURE 3** Evolution of $\Omega_e/\omega_{pe}=100$ PFA phase plots including Comptonization of external blackbody soft photons. Left panel is for low photon density and right panel is for high photon density, showing the rapid decline of the maximum Lorentz factors (Sugiyama these proceedings).

However, classical radiation damping using the Dirac term does not include Compton drag by external photons, which must be computed with Monte Carlo (MC) methods. We has recently implemented an approximate treatment of inverse Compton loss in the NN code by adding a time-averaged damping force $\mathbf{f}_D = -4\sigma_T g^2 \mathbf{v} U_s/3c$, where $\sigma_T$= Thomson cross-section and $U_s$=soft photon energy density, to represent the net effect of scattering by many isotropic soft photons [4]. Fig.3 gives sample results.

## INTERACTION WITH AMBIENT PLASMA

Astrophysically it is important to study the interaction of PFAs with the ambient environment, both to see how the ambient plasmas damp the acceleration and absorb the Poynting flux, and how the heated ambient plasma radiate. Fig.4 shows sample results of 3D runs. Some of our major findings include: (a) the maximum Lorentz factors achieved even with ambient plasma are comparable to the vacuum case of Fig.1, though the number of high-energy particles decreases with increasing ambient density. (b) a complex multi-phase plasma exists in the contact region where high-$\gamma$ ejecta plasma overruns low-$\gamma$ "swept-up" plasma. The swept-up plasma consists of 2 phases: a cold unaccelerated phase coexisting with a PF-accelerated phase whose Lorentz factors < those of the ejecta. In the e-ion case, both the "swept-up" electrons and ions exist in 2 distinct phases. The transition region is very broad (>>> the ion gyroradii, Fig.5), contrary to MHD results. (c) the swept-up ion Lorentz factors are << the electron Lorentz factors because ions are pulled only by the charge separation electric fields, while leptons are accelerated by the EM pulse directly. This disproves the conventional MHD assumption that the electron and ion bulk Lorentz factors are the same and the transition layer thickness is ~ ion gyroradius. (d) there is no evidence of any disruptive plasma instability (Weibel, 2-stream, and lower-hybrid-drift instabilities) at the contact interface. Such instabilities are suppressed to first order by the strong transverse EM field. (e) the Poynting flux decays via ponderomotive acceleration of the ejecta and ambient electrons, plus mode conversion to longitudinal plasma waves which are absorbed by Landau damping. Fig.6 shows that PFA preferentially accelerates the e+e- component in a mixture of e-ion and e+e- plasma.

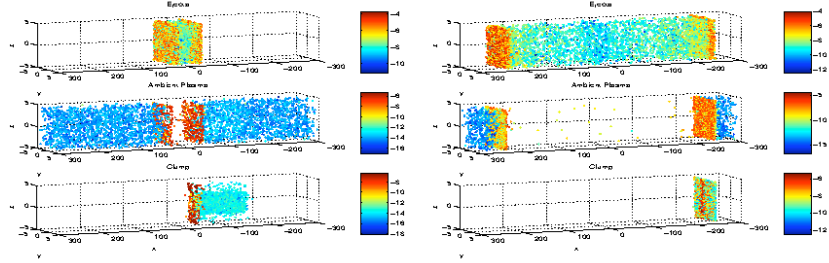

**FIGURE 4** 3D evolution of internal shocks when a warm e+e- ejecta slab with Lorentz factor of 100 at the front collides with e+e- plasmas upstream. The plasma on the left half of the grid is uniform but the plasma on the right has an additional high density "clump" to illustrate the effects of inhomogeneous shock. We plot 1/10000 of the particles and color code each particle with their instantaneous radiation power from shock acceleration. This example shows that the shock is not disrupted by inhomogeneity, and the transition layer is much thicker than gyroradii expected from MHD theory. We find that peak radiation power per particle is roughly the same for all three species: ejecta, upstream plasma and clump (from [8]).

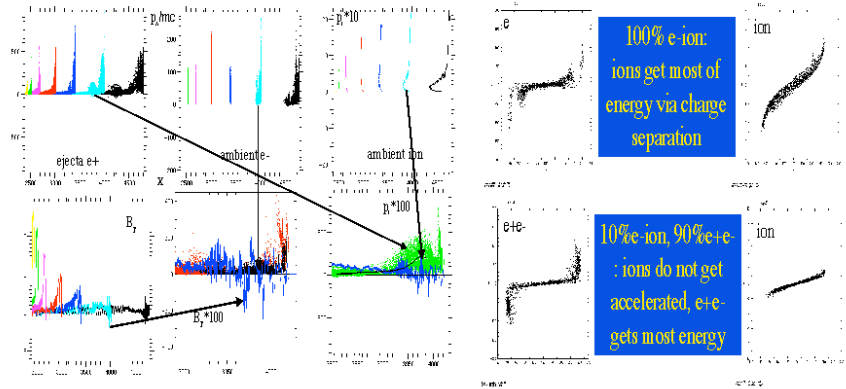

**FIGURE 5** (left) Time-lapse phase plots and B-field profiles for a slab e+e- 2.5D PFA running into (0.001x ejecta density) cold e-ion ambient plasma. $P_i$ is in units of $m_i c$. The acceleration of ejecta plasma stalls after it has swept up roughly equal mass of ambient plasma but the acceleration of ambient ions continues due to the pull by charge separation Ex. EM field decays by acceleration of swept-up plasma and mode conversion. Lower right panels show the blow-up details of the transition region [8].
**FIGURE 6** (r) PF acceleration of pure e-ion plasma compared to mixture of e-ion and e+e- plasma [8].

## ACKNOWLEDGMENTS

This work was partially supported by NASA NAG 5-9223 and NSF AST0406882